\documentclass{article}

\usepackage{arxiv}

\usepackage[utf8]{inputenc} 
\usepackage[T1]{fontenc}    
\usepackage{hyperref}       
\usepackage{url}            
\usepackage{booktabs}       
\usepackage{amsfonts}       
\usepackage{nicefrac}       
\usepackage{microtype}      
\usepackage{lipsum}
\usepackage{graphicx}
\graphicspath{ {./images/} }
\usepackage{braket}
\usepackage{amsmath}
\title{Geometric representation of adiabatic distributed-Bragg-reflectors and broadening the photonic bandgap}

\author{
 Shailja Sharma \\
 School of Physical Sciences\\ National Institute of Science Education and Research\\ HBNI, Jatni - 752050, Odisha, India\\
  \texttt{shailja.sharma@niser.ac.in} \\
   \And
 Abhishek Mondal \\
  School of Physical Sciences\\ National Institute of Science Education and Research\\ HBNI, Jatni - 752050, Odisha, India\\
  \texttt{abhishek.mondal@niser.ac.in} \\
  \And
 Ritwick Das* \\
  School of Physical Sciences\\ National Institute of Science Education and Research\\ HBNI, Jatni - 752050, Odisha, India\\
  \texttt{ritwick.das@niser.ac.in} \\
}

\begin{document}
\maketitle
\begin{abstract}
Adiabatic following has been an widely-employed technique for achieving near-complete population transfer in a `two-level' quantum mechanical system. The theoretical basis, however, could be generalized to a broad class of systems exhibiting SU(2) symmetry. In the present work, we present an analogy of population transfer dynamics of two level atomic system with that of light propagation in a classical `one-dimensional' photonic crystal, commonly known as distributed-Bragg-reflector (DBR). This formalism facilitates in adapting the idea of adiabatic following, more precisely the rapid adiabatic passage (RAP) which is usually encountered in a broad class of quantum-mechanical systems. We present a chirped DBR configuration in which the adiabatic constraints are satisfied by virtue of optimally chirping the DBR. The reflection spectrum of the configuration exhibit broadening of photonic bandgap (PBG) in addition to a varying degree of suppression of sharp reflection peaks in the transmission band. The intermodal coupling between counter-propagating modes as well as their phase-mismatch, for the DBR configuration, exhibits a longitudinal variation which is usually observed in `Allen-Eberly' scheme of adiabatic population transfer in \emph{two-level} atomic systems.
\end{abstract}


\section{Introduction}
The propagation characteristics of electromagnetic ($em$) waves in a periodically stratified dielectric medium closely resemble the features exhibited by the matter waves in crystalline solids \cite{PhysRevLett.58.2059,Optical}. In crystals, the periodic \emph{Coulomb}-potential leads to formation of continuous energy bands separated by forbidden energy spectrum, also  known as bandgaps. In resemblance, analogous periodic photonic architectures, commonly termed as \emph{photonic crystals} (PCs), lead to formation of $em$ transmission \emph{bands} enclosed by forbidden frequency spectrum known as photonic bandgap (PBG) \cite{Joannopoulos,Yariv:77}. Over the last four decades, PCs has evolved as backbone of many technological advancements which essentially hinge upon manipulating the spatial and spectral characteristics of light beam \cite{pochi}. In one-dimension, the PCs are also known as distributed-Bragg-reflector (DBR) and they have turned out to be primary ingredients for devising reflectors/anti-reflectors, spectral and spatial filtering elements and creating efficient sensing platforms which include those involving plasmonic interactions \cite{PhysRevB.76.165415,Shukla:18,Mitsuteru,Maigyte,Shen,PhysRevE.71.066604}. Bragg reflection based waveguiding configurations have been employed as polarization selection device in miniaturized optical sources \cite{West,Simova}. Bragg-reflection-waveguides (BRWs) with a vacuum core have been forecasted as plausible hosts for future optical particle accelerators driven by extremely high power lasers \cite{PhysRevE.70.016505}. On the other hand, the possibility of field confinement and therefore, reducing the $em$ interaction volume to sub-wavelength scale has lead to the formation of stable spatial-solitons in nonlinear BRWs \cite{Wchter1992NonlinearBR}. Recently, the bulk modes of a BRWs or optical surface states in DBR have been found to be promising candidates for optical parametric processes and higher harmonic generation \cite{PhysRevA.86.023819,Afinogenov,kivshar}. As mentioned before, a DBR is primarily characterized by PBG whose magnitude is primarily dictated by the refractive index contrast of the DBR constituents and the location in a spectral band is governed by the thickness of constituent layers \cite{pochi}. Therefore, for a given pair of material forming the DBR, the magnitude of PBG is unique and fixed. Within the limit of optical transparency for the constituent DBR materials, we present a formalism to broaden the PBG (and suppress the transmission band) by utilizing the concept of adiabatic coupling to a backward propagating mode from a forward propagating mode. Therefore, we draw an equivalence of the coupled-wave equations in a DBR with that encountered while describing the dynamics of quantum \emph{two-level} atomic systems. Subsequently, we propose plausible DBR configurations for adopting the adiabatic following in such systems which leads to broadening of PBG spectrum. The formalism could be translated to any spectral band and PBG broadening is limited by the restriction imposed by material transparency only. Interestingly, this idea provides an viable platform to tailor the backscattered phase from such systems as well. 

Adiabatic following, also known as rapid adiabatic passage (RAP) has been an well-established technique for realizing near $100\%$ population transfer in a \emph{two-level} atomic systems using optimum ultrashort ($\leq~0.5~ps$) pulses \cite{eberly}. In fact, it is quite straightforward to ascertain that the RAP mechanism could be adopted in any equivalent system exhibiting \emph{special unitary(2)} or \emph{SU(2)} symmetry. For example, RAP provides an efficient platform for ultrashort pulse frequency conversion as well as tailoring spatial characteristics of optical beams in optical nonlinear medium \cite{suchowski,Karnieli:18,Karnieli,PhysRevA.101.033807}.
\section{Adiabatic-following in DBR}
A periodic variation (with periodicity $\Lambda$) in dielectric constant ($\epsilon$) along $z$-direction in a medium could be expressed as \cite{yariv}
\begin{equation}\label{eq:equation1}
    \epsilon (x,y,z) = \epsilon (x,y) + \Delta \epsilon (x,y,z)
\end{equation}

where $\Delta \epsilon (x,y,z)$ defines the periodic modulation along $z$. This dielectric perturbation gives rise to the possibility of intermodal interactions. Let us consider any two modes, namely $\ket{p}$ and $\ket{q}$ propagating through the medium described by dielectric function in equation \eqref{eq:equation1}. Assuming the modes are propagating along $z$-direction, the $em$-field corresponding to the modes (in the paraxial limit) are given by $\ket{p}~=~E_p(x,y)A_p(z)e^{i(\omega t - \beta_p z)}$ and $\ket{q}~=~E_q(x,y)A_q(z)e^{i(\omega t - \beta_q z)}$. Under the slowly varying approximation, the evolution of $z$-dependent mode-amplitude ($A_p$) of the $p^{th}$-mode due to the presence of $q^{th}$-mode (with amplitude $A_q$) is given by \cite{yariv}
\begin{equation}\label{eq:equation2}
\frac{d A_q}{dz} = -i \frac{\beta_q}{|\beta_q|} \sum_{p} \sum_m {\kappa_{pq}}^{(m)} A_p e^{-i(\beta_p - \beta_q -m\frac{2\pi}{\Lambda}) z}
\end{equation}
where $\beta_p$ and  $\beta_q$ are the normal components of wavevector $k_p$ and $k_q$ respectively. ${\kappa_{pq}}^{(m)}$ defines the magnitude of coupling coefficient which couples the modes through the $m^{th}$ Fourier component of dielectric function equation \eqref{eq:equation1}. The $m^{th}$ coupling coefficient is expressed as  
\begin{equation}\label{eq:equation3}
{\kappa_{pq}}^{(m)} = \frac{\omega}{4} \int \int {E^*}_p(x,y) \epsilon_m(x,y) {E}_q(x,y) dx dy
\end{equation}
where $\epsilon_m$ is the $m^{th}$ component of Fourier-series expansion of $\epsilon$ (see equation \eqref{eq:equation1}). The coupling between the interacting modes is maximum when the longitudinal phase-matching condition is exactly satisfied, 
\begin{equation}\label{eq:equation4}
 \beta_p - \beta_q -m\frac{2\pi}{\Lambda} = 0
\end{equation}
In case of DBR, the coupling between a forward propagating mode ($A_p \equiv A_i$) and a backward propagating mode ($A_q \equiv A_r$) is of interest. In absence of any other intermodal interactions, the forward-backward mode coupling (assuming $m=1$) would be governed by \cite{yariv,GARG2013148,Thyagarajan},
\begin{equation} \label{eq:equation5}
  \frac{dA_r}{dz} = - i \frac{\beta_r}{|\beta_r|} {\kappa} A_i e^{i \Delta \beta z}  
\end{equation}
\begin{equation} \label{eq:equation6}
  \frac{dA_i}{dz} = - i \frac{\beta_i}{|\beta_i|} {\kappa^{*}} A_r e^{-i \Delta \beta z}  
\end{equation}
where we have dropped the indices from $\kappa$ and $\Delta \beta = \beta_r - \beta_i -\frac{2\pi}{\Lambda}$. In case the DBR constituent materials are isotropic, the dielectric function (equation \eqref{eq:equation1}) is a purely scalar and consequently, transverse-electric ($TE$) to transverse-magnetic ($TM$) mode-coupling (and vice-versa) is forbidden. It is worth noting that the phase-matching condition for coupling the forward propagating mode ($\ket{i}$) to a backward propagating mode ($\ket{r}$) or vice-versa is contra-directional in nature and therefore, Eq. \eqref{eq:equation4} could be expressed as $\Delta \beta = 2 \beta -  \frac{2\pi}{\Lambda} = 0$ where $\beta_i=-\beta_r=\beta$ \cite{yariv}. As shown in Fig.1, the phase-matching condition for the contra-directional coupling process is modified to $\Delta \beta = 2 \beta \cos{\theta} -  \frac{2\pi}{\Lambda} = 0$ in case of oblique incidence ($\theta$ with respect to $z$-axis),. For a simple case when the two layers of the DBR (with refractive indices $n_1$ and $n_2$) share the same thickness \emph{i.e.} $d_1~=~d_2$, the coupling coefficient takes a simplified form given by \cite{yariv}, 
\begin{equation}\label{eq:equation7}
    \kappa_{TE} = \frac{i}{\lambda \cos{\theta}} \frac{\sqrt{2} ({n_1}^2 - {n_2}^2)}{\sqrt{({n_2}^2 + {n_1}^2)}} 
\end{equation}
\begin{equation}\label{eq:equation8}
    \kappa_{TM} = \frac{i}{\lambda \cos{\theta}} \frac{\sqrt{2} ({n_1}^2 - {n_2}^2)}{\sqrt{({n_2}^2 + {n_1}^2)}} \cos{2\theta}  
\end{equation}
The solution of equations \eqref{eq:equation5} and \eqref{eq:equation6} for \emph{close-to-perfect} phase-matched ($\Delta\beta~\approx~0$) situation leads to a strong coupling from a forward propagating mode ($\ket{i}$) to a backward propagating mode ($\ket{r}$) for broad range of frequencies. This is realized for all the frequencies within the PBG and consequently, we obtain a strong reflection band. It is apparent from equations \eqref{eq:equation7} (or \eqref{eq:equation8}) that the contrast in refractive index (or dielectric constant) is the primary factor determining the strength of coupling and hence, the sharpness of band edges. For a given choice of DBR constituents, $\vert n^{2}_1 - n^{2}_2 \vert$ is fixed and consequently, the width of PBG remains unchanged. 
\begin{figure}[htbp]
\centering\includegraphics[width=8cm]{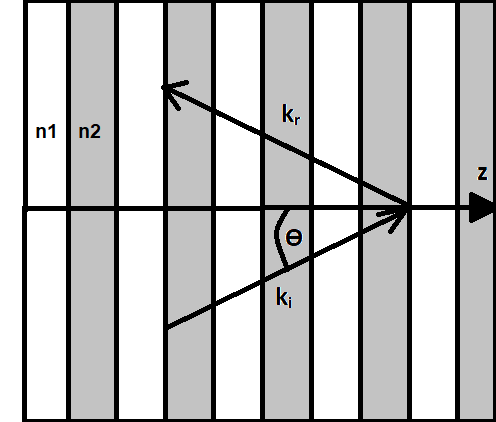}
\caption{Oblique incidence of $em$-wave on distributed-Bragg-reflector (DBR). $z$-direction represents the optical axis}
\label{Figure1}
\end{figure}
\subsection{Adiabatic phase-matching}
We consider the following transformation to a rotating frame given by
\begin{equation}\label{eq:equation9}
    A_r = \tilde{a_r} e^{i/2[\Delta\beta(0) z - {\int_0}^z q(\tilde{z}) d\tilde{z}]}
\end{equation}
\begin{equation}\label{eq:equation10}
  A_i   = \tilde{a_i} e^{-i/2[\Delta\beta(0) z - {\int_0}^z q(\tilde{z}) d\tilde{z}]}
\end{equation}
where $\Delta \beta (z) = \Delta\beta(0) z - {\int_0}^z q(\tilde{z}) d\tilde{z}$ and $\Delta\beta(0)$ is the phase-mismatch at $z=0$, $q(z)$ is determined by the dielectric constant modulation $\Delta\epsilon$ (as per Eq. \eqref{eq:equation1}), and $\Delta k = \frac{q(z) - \Delta \beta(0)}{2}$. The complex amplitudes $\tilde{a_i}$ and $\tilde{a_r}$ represent the power contained in the forward and backward propagating modes in the rotated reference frame. It is worthwhile to note that $\vert\tilde{a_i}\vert^2 = \vert{A_i}\vert^2$ and $\vert\tilde{a_r}\vert^2 = \vert{A_r}\vert^2$. The substitution of equation \eqref{eq:equation9} and \eqref{eq:equation10} into equations \eqref{eq:equation5} and \eqref{eq:equation6} yields
\begin{equation} \label{eq:equation11}
-i\frac{d}{dz} \begin{pmatrix} \tilde{a_r}\\ \tilde{a_i}\\ \end{pmatrix} = \begin{pmatrix} \Delta k & \kappa\\
-\kappa^{*} & -\Delta k \\ \end{pmatrix}
\begin{pmatrix} \tilde{a_r}\\ \tilde{a_i}\\ \end{pmatrix}
\end{equation}
which is an equivalent representation of time-dependent Schrödinger equation in optics. In fact, if we define a state $\big| \Psi \big> = \bigg(\begin{matrix} \tilde{a_r}\\
\tilde{a_i}\end{matrix}\bigg)$ then $-i\frac{d}{dz}\ket{\Psi} = \hat{H} \ket{\Psi}$ where time-dependence is replaced by the $z$-dependence and the Hamiltonian $\hat{H} = \vec{\sigma} . \vec{B}$. Here, $\vec{\sigma} = (\sigma_x , \sigma_y, \sigma_z)$ are the Pauli-spin matrices and $\vec{B} = (\kappa, 0, \Delta k)$ represents a fictitious magnetic field. The Hamiltonian has a close resemblance with that representing the dynamics of a spin-$1/2$ particle in an external magnetic field where $\ket{i}$, $\ket{r}$ are the equivalent to the spin-up ($\ket{\uparrow}$) and spin-down ($\ket{\downarrow}$) states\cite{Feynman1957GeometricalRO}. In order to represent a particular pseudo-spin state, the equivalent Stokes parameters could be given by $S_{j} = \big< \sigma_{j} \big>$ ($j\equiv x,y,z$) and the evolution of the state would be dictated by $\frac{\partial{S}_{j}}{\partial z} = -i \big<\big[ \sigma_{j}, \hat{H} \big]\big> $ . This, subsequently, leads to
\begin{equation} \label{eq:equation12}
   \begin{matrix} S_x = \Tilde{a_i} {\Tilde{a_r}}^* + \Tilde{a_r}{\Tilde{a_i}}^*\\
    S_y = -i[\Tilde{a_i} {\Tilde{a_r}}^* - \Tilde{a_r} {\Tilde{a_i}}^*]\\
    S_z = {| \Tilde{a_r}|}^2 - {| \Tilde{a_i}|}^2
    \end{matrix}
\end{equation}
and
\begin{equation}\label{eq:equation13}
    \Dot{\Vec{S}} = \Vec{B} \times \Vec{S}
\end{equation}
where $\Dot{\vec{S}}\equiv\frac{\partial \Vec{S}}{\partial z}$). The states and their evolution described by equations \eqref{eq:equation12} and \eqref{eq:equation13} determine the evolution dynamics of the forward and backward propagating modes. In other words, equations \eqref{eq:equation12} and \eqref{eq:equation13} could be mapped onto an equivalent Bloch-sphere where $\ket{i}$ (spin-up) and $\ket{r}$ (spin-down) states are located the \emph{south}-pole and \emph{north}-pole respectively. Any point of the DBR Bloch-sphere represents a particular superposition of eigenstates $\ket{i}$ and $\ket{r}$. \emph{Sec. 2.2} discusses the geometric representation in detail for specific cases of evolution of these states. The fictitious magnetic-field ($\vec{B}$) constitutes a parameter space for exerting a \emph{torque} on the system which leads to precession of the state vector ($\Vec{S}$) about $\Vec{B}$ with a frequency $|\Vec{B}| = \sqrt{{|\kappa|}^2 + {\Delta k}^2}$. It is worthwhile to point out that the value $S_z$ (equation \eqref{eq:equation12}) determines the mode-conversion (from the forward-to-backward or vice-versa) factor. The state $\vec{S}=[0,0,-1]$ represents a situation where the incident beam contains all the optical power $i.e.~\tilde{a}_i = 1$ and the state $\vec{S}=[0,0,1]$ represents the case when all the optical power is present in the reflected beam ($\tilde{a}_r = 1$). The conversion efficiency to the reflected beam is expressed as $\eta=\frac{S_z+1}{2}$ as shown in Fig. \ref{Figure5}. The dynamics, here, is primarily controlled by the parameters $\Delta k$ and $\kappa$ which are essentially equivalent to factors \emph{detuning} ($\Delta$) and \emph{Rabi-frequency} ($\Omega$) respectively which are commonly encountered in a \emph{two-level} atomic system. In a perfectly phase-matched case ($\Delta k = 0$), which is satisfied for the central PBG frequency, a gradual semi-circular rotation of the state-vector ($\vec{S}$) from south-pole to the north-pole takes place on DBR Bloch-sphere which implies a complete transfer of optical power from the incident beam to the reflected beam in the DBR. In order not to obscure the salient features of the equivalent dynamics, we have ignored the role played by absorption loss ($\alpha$) in the dielectric materials. However, it would be worthwhile to point out that the role of absorption loss is equivalent to that played by semi-phenomenological decay constant ($T_2$) in a \emph{two-level} atomic system \cite{eberly}. For $\Delta k \neq 0$, $\vec{S}$ traces a trajectory on the DBR Bloch-sphere about the axis $\hat{n} = \frac{\kappa \hat{x}+\Delta k \hat{z}}{\sqrt{\kappa^2+\Delta k^2}}$. This implies that $\Delta k \neq 0$ leads to transportation of state-vector $\vec{S}$ to a point in northern-hemisphere. This point may be near the north-pole but it will never be the north-pole of the equivalent DBR Bloch-sphere. Nevertheless, $S_z$ still positive and the reflectivity is large (not maximum). The frequencies for which $S_z$ remains positive forms a band of frequencies constituting the PBG. However, a sufficiently large $\Delta k$ results in a $\hat{n}$ such that $\vec{S}$ is transported to a point which lies on the southern hemisphere of DBR Bloch-sphere (assuming $\vec{S}$ starts precessing about $\vec{B}$ from south-pole). Therefore, $S_z$ would be negative after propagation through the DBR. This corresponds to frequencies outside the PBG (or within the transmission band) where reflectivity is much smaller than the PBG. In fact, the \emph{equator} of DBR Bloch-sphere distinguishes the PBG (northern hemisphere) from the transmission (pass) band (southern hemisphere) with respect to the trajectory of $\vec{S}$.

It is apparent that $\Delta k = 0$ along the entire DBR length for the central PBG frequency and $\Delta k \neq 0$ for all other frequencies. The aforementioned analogy to a \emph{two-level} atomic system, thus, allows us to adopt the Rapid Adiabatic Passage (RAP) mechanism for broadening the PBG spectrum and suppressing the transmission (pass) bands in the DBR. Consequently, we propose DBR configurations which leads to \emph{slow} variation in $\Delta k$ from a large negative to a large positive value along the DBR length such that the sweep always remains much smaller than the coupling ($\kappa$). This would ensure nearly complete transfer of optical power from a forward propagating to a backward propagating mode across a frequency band extending much beyond the conventional PBG of a DBR. The RAP condition is expressed as \cite{eberly,Bahar:18}
\begin{equation}\label{eq:equation14}
    \kappa \Dot{\Delta k} - \Delta k \Dot{\kappa} << {({|\kappa|}^2 + {\Delta k}^2)}^{3/2}
\end{equation}
which essentially implies that the \emph{rate} of change of polar angle $\Phi = {tan}^{-1}(\frac{\kappa}{\Delta k})$ (angle between $\Vec{B}$ and $z$ axis) during the evolution is much slower with the frequency of rotation ($=\sqrt{\kappa^2+\Delta k^2}$) for $\vec{S}$ about $\vec{B}$. If $\kappa$ is assumed to be constant (along $z$), the adiabaticity condition appears as $\frac{d\Delta k}{dz} << \frac{{(\kappa^2 +{\Delta k}^2)}^{3/2}}{\kappa}$. Also, in order to achieve near-complete optical power transfer, it is essential that the two states (modes) $\ket{i}$ and $\ket{r}$ are decoupled at the entry ($z=0$) and exit ($z=L$) faces of the DBR. Alternately, this is mathematically expressed as $\vert\frac{\Delta\beta}{\kappa}\vert~>>~2$ at $z=0,L$ which is equivalent to satisfying the condition of \emph{autoresonance} in `two-wave' interaction system \cite{Markov:18,Yaakobi:13}. In this case, \emph{autoresonance} essentially ensures that the counter-propagating modes remain phase-locked when the parameters of the Hamiltonian undergo an adiabatic change. This manifests into a near-complete transfer of optical power from the forward to the backward propagating mode. 

\subsection{Chirped DBR configuration for adiabatic mode-conversion}
\begin{figure}[htbp]
\centering\includegraphics[width=14cm]{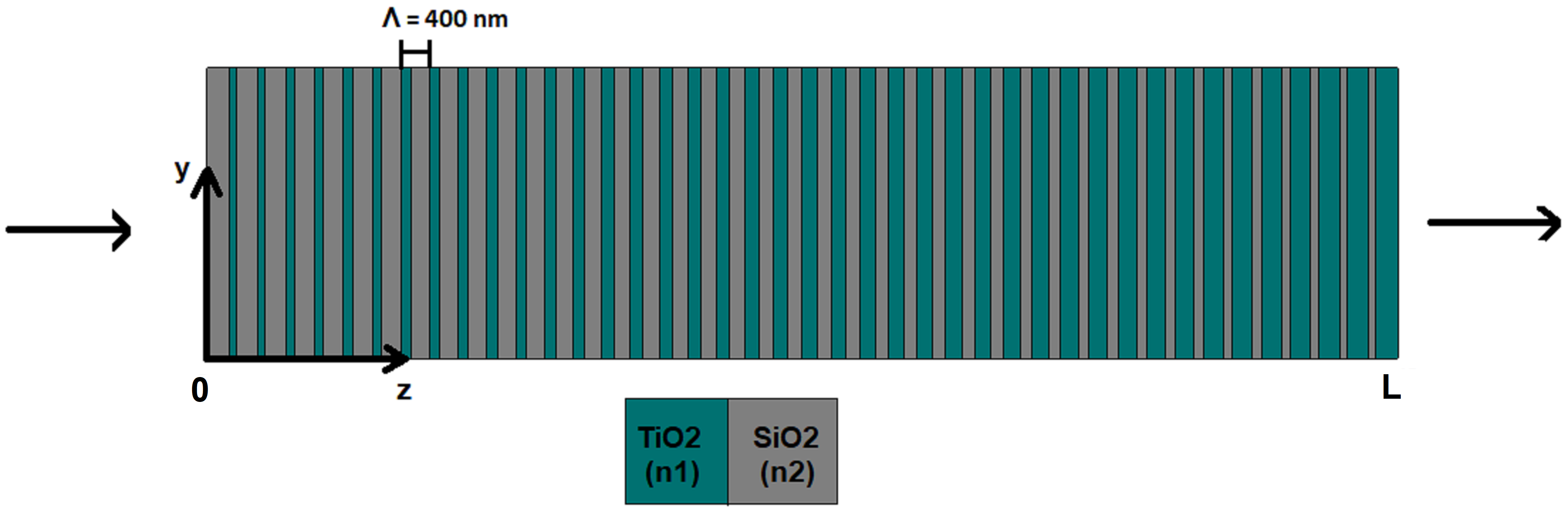}
\caption{A schematic to describe the geometry of a chirped-DBR}
\label{Figure2}
\end{figure}
In order to adopt an adiabatic coupling scheme, we consider a DBR configuration with a linear chirp in the duty cycle of each unit cell \emph{i.e.} thickness $d_{1M} = d_1+ M\delta$ and $d_{2M} = \Lambda - d_1 - M\delta$ define the thickness of layers $A$ and $B$ respectively in $M^{th}$ unit cell as shown in Fig. \ref{Figure2}. The unit cell period ($\Lambda$), however remains unchanged. Here, $M = 0,1,2,3,...,(N-1)$ where $N$ is the total number of unit cells in the DBR. This leads to a longitudinal variation in $\Delta k$ through a monotonic change in average refractive index of an \emph{unit cell} which could be defined as $\Bar{n} = \sqrt{\frac{d_{1M}{n_1}^2 +  d_{2M}{n_2}^2}{\Lambda}}$. It is worthwhile to note that the variation of $\Bar{n}$ also manifests in the form $z$-dependence of $\kappa$. Using equations \eqref{eq:equation4}, \eqref{eq:equation7} and \eqref{eq:equation8}, the variation in $\Delta k$ and $\kappa$ for normal incidence ($\theta = 0$) would be expressed as \cite{yariv}

\begin{equation}\label{eq:equation17}
    \Delta k = \frac{\Delta\beta}{2} = \frac{2\pi \Bar{n}}{\lambda} - \frac{\pi}{\Lambda}
\end{equation}
\begin{equation}\label{eq:equation18}
    \kappa =  \frac{i(1-\cos{(\frac{2\pi d_{1M}}{\Lambda}}))}{2\lambda} \frac{({n_1}^2 - {n_2}^2)}{\Bar{n}}
\end{equation}\\

\begin{figure}[htbp]
\centering\includegraphics[width= \linewidth]{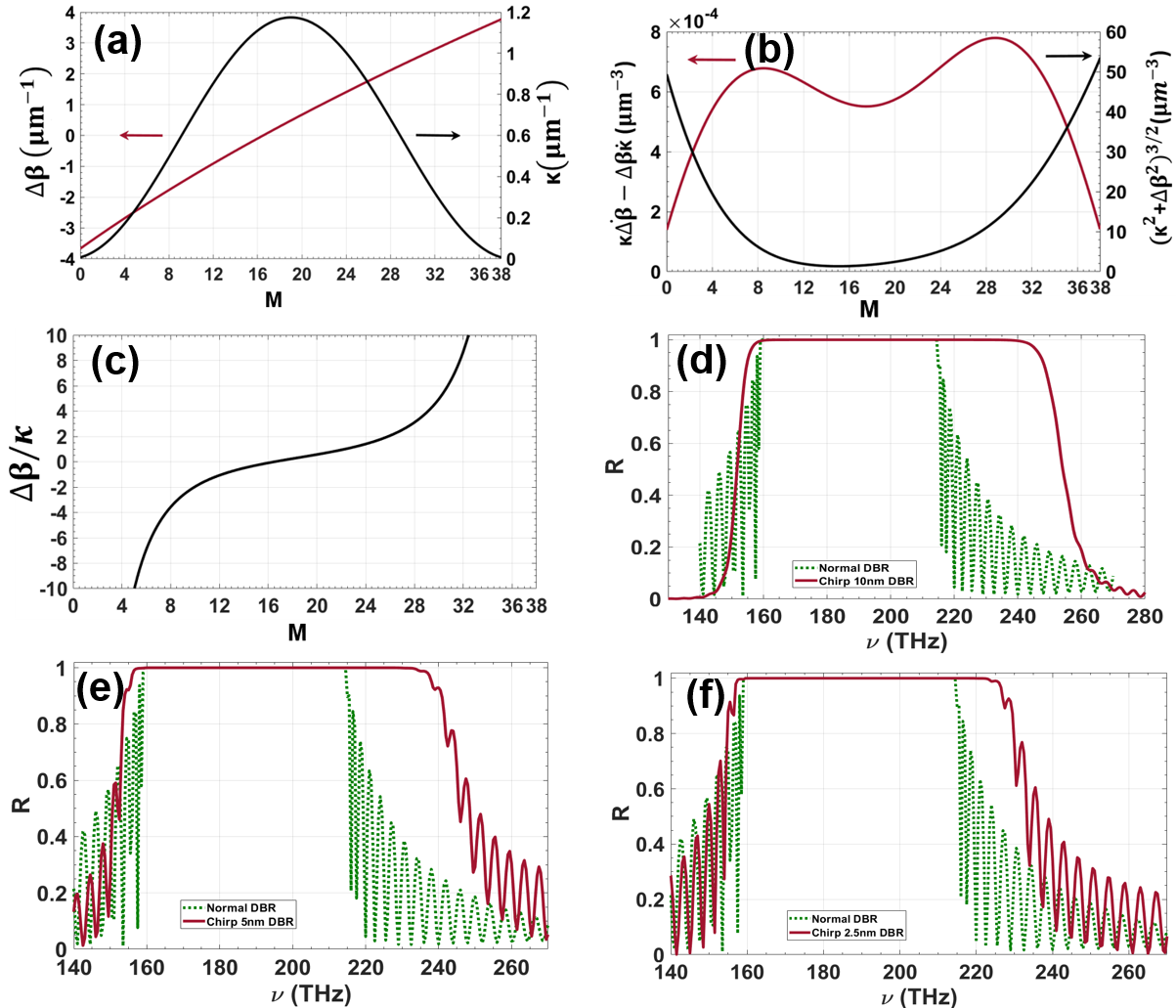}
\caption{a) shows the variation of $\Delta \beta$ and $\kappa$ in $M^{th}$ unit cell. b) shows the variation of LHS and RHS of the inequality given in Eq. \eqref{eq:equation14} in $M^{th}$ unit cell. c) shows the variation of $\frac{\Delta \beta}{\kappa}$ as a function of unit cell no. ($M$) for depicting a significant fraction of DBR length satisfies the \emph{(condition for auto-resonance)}.  d), e) and f) shows a comparison between the reflection spectrum of a normal-DBR (dashed green line) of $d_1 = d_2 = 200~nm$ with that for C-DBRs (solid red line) having $\delta=~10~nm$, $\delta=~5~nm$ \& $\delta=~2.5~nm$ respectively.}
\label{Figure3}
\end{figure}

For an arbitrarily chosen chirp-length of $\delta = 10~nm$, $d_1 = 10~nm$ and $N=39$, the variation in $\Delta k$ and $\kappa$ for the chirped-DBR (C-DBR) is shown in Fig. \ref{Figure3}(a). It is apparent that $\Delta\beta$ ($=2\Delta k$) varies symmetrically from a large negative (at $z=0$ or $M=0$) to a large positive value (at $z=L$ or $M=38$). The coupling coefficient ($\kappa$), on the other hand, reaches a maximum at the center of C-DBR geometry ($M~=~20$) and negligibly small at $z=0,L$. Figure \ref{Figure3}(b) shows the variation of $\kappa \frac{d\Delta \beta}{dz} - \Delta \beta \frac{d\kappa}{dz}$ in $M^{th}$ unit cell of C-DBR. It is apparent that this is much smaller than $(\kappa^2+\Delta \beta^2)^{3/2}$ at any point within the C-DBR. Therefore, the adiabaticity condition described by equation \eqref{eq:equation14} is completely satisfied in case of C-DBR. It is interesting to note that the auto-resonant condition $i.e.~\big\vert\frac{\Delta \beta}{\kappa}\big\vert < 2$ is satisfied in a significant fraction of the C-DBR ($M\approx10$ to $M\approx28$) as shown in Fig. \ref{Figure3}(c). This essentially implies that the counter-propagating modes (at a certain frequency) would be phase-locked in the entire interaction region if they are phase-matched ($\Delta \beta = 0$) in any one \emph{unit cell} of C-DBR. 

The impact of closely satisfying the adiabatic constraints have a profound impact on reflection spectrum shown in Fig. \ref{Figure3}(d). In order to obtain the reflection spectrum, we consider a cross-section as shown in Fig. \ref{Figure2} where $N=39$ is considered. A broadband plane-wave is incident on the C-DBR from $z = 0$. The reflection spectrum was obtained using finite element method (FEM) (wave-optics module, COMSOL Multiphysics). The periodic boundary condition is imposed along the transverse direction and a mesh-size of $5~nm$ is considered for the simulation. Also, we ignore the material dispersion in the present case and assumed $n_1 = 2.5$ ($A\equiv TiO_2$) and $n_2 = 1.5$ ($B\equiv SiO_2$) across the entire spectrum. In order to compare, the reflection spectrum for a normal DBR (no-chirping and $\Lambda = 400~nm$) is also shown in \ref{Figure3}(d). It is evident that there is $\approx~240~nm$ increase in PBG for C-DBR as compared to the normal DBR. Also, the reflectivity drop at the band-edges is relatively smooth with complete suppression of small reflection resonances outside the PBG. In fact, any alteration in structural parameters has an impact $\Delta \beta$ and $\kappa$ and consequently, the adiabatic constraints are not satisfied closely. This is expected to reduce the PBG in reflection spectrum along with appearance of sharp transmission resonance (outside PBG). For example, when the chirp-length changes to $\delta = 5~nm$ with $d_1 = 100~nm$ and $N=39$, the reflection spectrum is shown in Fig. \ref{Figure3}(e) where the PBG shrinks to $\approx~182~nm$. It is also worth noting that the reflection spectrum is accompanied by oscillating side-bands with sharp transmission resonances on both ends of the PBG. On reducing the chirp length further to $\delta = 2.5~nm$ ($d_1=150~nm$ and $N=39$), PBG for the C-DBR shrinks to $100~nm$ and discernibly sharper transmission resonances on both ends of PBG which is similar to that exhibited by normal DBR. The underlying cause behind this observation could be traced to the variation of $\kappa$ (see equation \eqref{eq:equation18}). By reducing the chirp length (without changing the DBR length $L$), the minimum value of $\kappa$ (at $z=0$) and $z=L$) increases. Therefore, the forward and backward propagating modes are not completely decoupled at the ends of C-DBR when $\delta = 5~nm$ and $\delta = 2.5~nm$. However, reducing the chirp has a weak impact on $\frac{d\Delta \beta}{dz}$. Overall, adiabaticity conditions are partially satisfied for smaller chirp length and consequently, we obtain a smaller PBG.

\begin{figure}[htbp]
\centering\includegraphics[width= \linewidth]{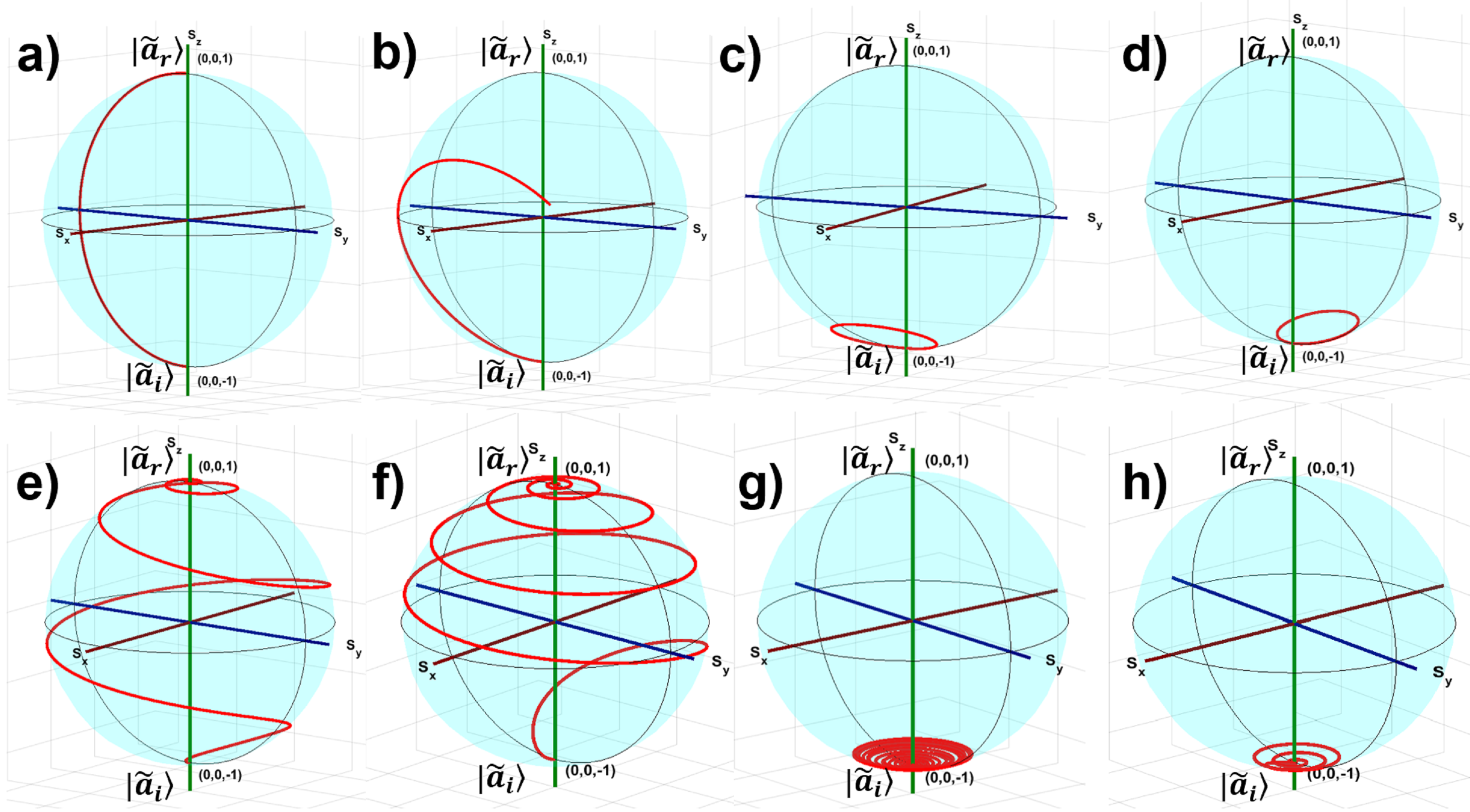}
\caption{Shows the evolution of state vector $\Vec{S}$ on the Bloch sphere for the normal DBR at wavelengths a) $\lambda_c = 1650~nm$ (central wavelength) b) $\lambda = 1500~nm$ c) $\lambda = 1000~nm $ d) $\lambda = 2500~nm$ and C-DBR at wavelengths (e) $\lambda_c = 1650~nm$ (central wavelength) (f) $\lambda = 1500~nm$ (g) $\lambda = 1000~nm$ and (h) $\lambda = 2500~nm$ respectively.}.
\label{Figure4}
\end{figure}
\subsection{Geometric representation of propagation characteristics in DBR}
\begin{figure}[htbp]
\centering\includegraphics[width= \linewidth]{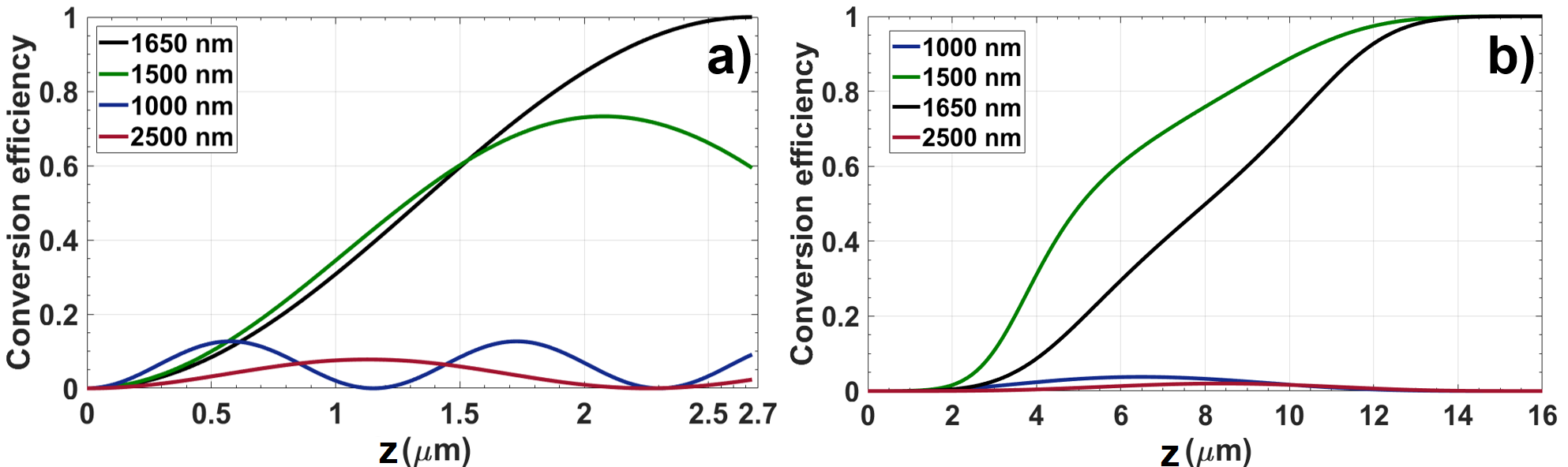}
\caption{a) and b) shows the variation of conversion efficiency as a function of propagation distance for normal DBR and C-DBR at wavelengths used in Fig. \ref{Figure4}. The black solid line represents the evolution of the central wavelength of the PBG, the green solid lines for wavelength that lie closer to the band-edges. The blue/purple solid lines correspond to wavelengths that lie outside the PBG}.
\label{Figure5}
\end{figure}

The analogy between a two-level atomic system and propagation dynamics in a DBR is apparent from Eqs. \eqref{eq:equation12} and \eqref{eq:equation13} where the time-evolution has been replaced by evolution along $z$-axis. The ground-state and excited state populations are analogues to complex amplitudes $\tilde{a_i}$ and $\tilde{a_r}$ respectively. The parameters determining the geometrical path $\Delta$ (frequency detuning) and $\Omega$ (Rabi frequency) are replaced by $\Delta k$ (phase-mismatch) and $\kappa$ (coupling coefficient). Therefore, the dynamical trajectory on the Bloch-sphere for a perfectly phase-matched ($\Delta k = 0$) interaction in the DBR should resemble an on-resonance interaction in two-level atomic system. Consequently, a $\pi$-pulse ($\kappa.z = \pi$) at the resonance is equivalent to exactly satisfying the Bragg's condition which manifests into complete conversion of $\tilde{a_i}$ into $\tilde{a_r}$ resulting in reflectivity $R = 1$. On an equivalent Bloch-sphere, Fig. \ref{Figure4}(a) represent such an evolution where the state-vector ($\vec{S}$) traverses from south-pole to the north-pole along the great circle (red-curve). Here, the DBR has periodicity $\Lambda = 400~nm$, $d_1 = d_2 = \frac{\Lambda}{2}$ and the PBG central wavelength is $\lambda_c = 1650~nm$ (where $\Delta k = 0$). Any non-zero $\Delta k$ (analogous to detuned interaction) would mean a smaller mode-conversion. In case of frequencies within the PBG (say $\lambda = 1500~nm$), the dynamical trajectory would result in termination of the state-vector at some point on the surface of northern-hemisphere of the DBR Bloch sphere as shown in Fig. \ref{Figure4}(b). Interestingly, for frequencies outside the PBG ($\vert\Delta k\vert >> 0$) or in the transmission band (say $\lambda = 1000~nm$ or $\lambda = 2500~nm$), the dynamical trajectory traced by $\vec{S}$ terminates within the southern hemisphere which could be observed in Figs. 4(c) and (d) respectively. The conversion efficiency ($\eta = \frac{S_z+1}{2}$) as a function of propagating distance ($z$) for all the aforementioned frequencies in normal DBR is shown in Fig.\ref{Figure5}(a). It could be noted that $\eta < 1$ for all wavelengths except central PBG wavelength ($\lambda_c = 1650~nm$) which is consistent with the description in Fig. \ref{Figure4}. 

The geometric representation for a C-DBR is shown in Fig. \ref{Figure4} for frequencies within the PBG (Fig. \ref{Figure4}e,f) as well as outside the PBG (Fig. \ref{Figure4}g,h). Since, the parameters $\Delta k$ and $\kappa$ exhibit a \emph{slow} longitudinal variation, $\vec{S}$ always remain perpendicular to $\vec{B}$ at each $z$. Therefore, $\vec{S}$ follows a spiralling trajectory after beginning its journey from the south-pole. Figure \ref{Figure4}(e) represents a dynamical trajectory of $\vec{S}$ (at $\lambda_c=1650~nm$) for the variation in $\Delta k$ (or $\Delta\beta$) shown in Fig. \ref{Figure3}(a). In this case, $\vec{S}$ traverses from the south-pole to the north-pole through a spiralling trajectory and this results in complete complete conversion of $\ket{i}$ to $\ket{r}$ as shown in Fig. 5(b). For another wavelength $\lambda = 1500~nm$ which is within the PBG, the dynamical trajectory appears similar but it exhibits asymmetrically distributed spirals (with respect to the equator) while travelling from the south-pole to the north-pole of DBR Bloch sphere (see Fig. \ref{Figure4}(b). In this case, the conversion efficiency reaches unity through a different path as shown in Fig. \ref{Figure5}(b). On the other hand, $\vec{S}$ remains within the southern hemisphere for wavelength outside the PBG (see Figs. \ref{Figure4}(g) and (h)). 

\begin{figure}[htbp]
\centering\includegraphics[width= \linewidth]{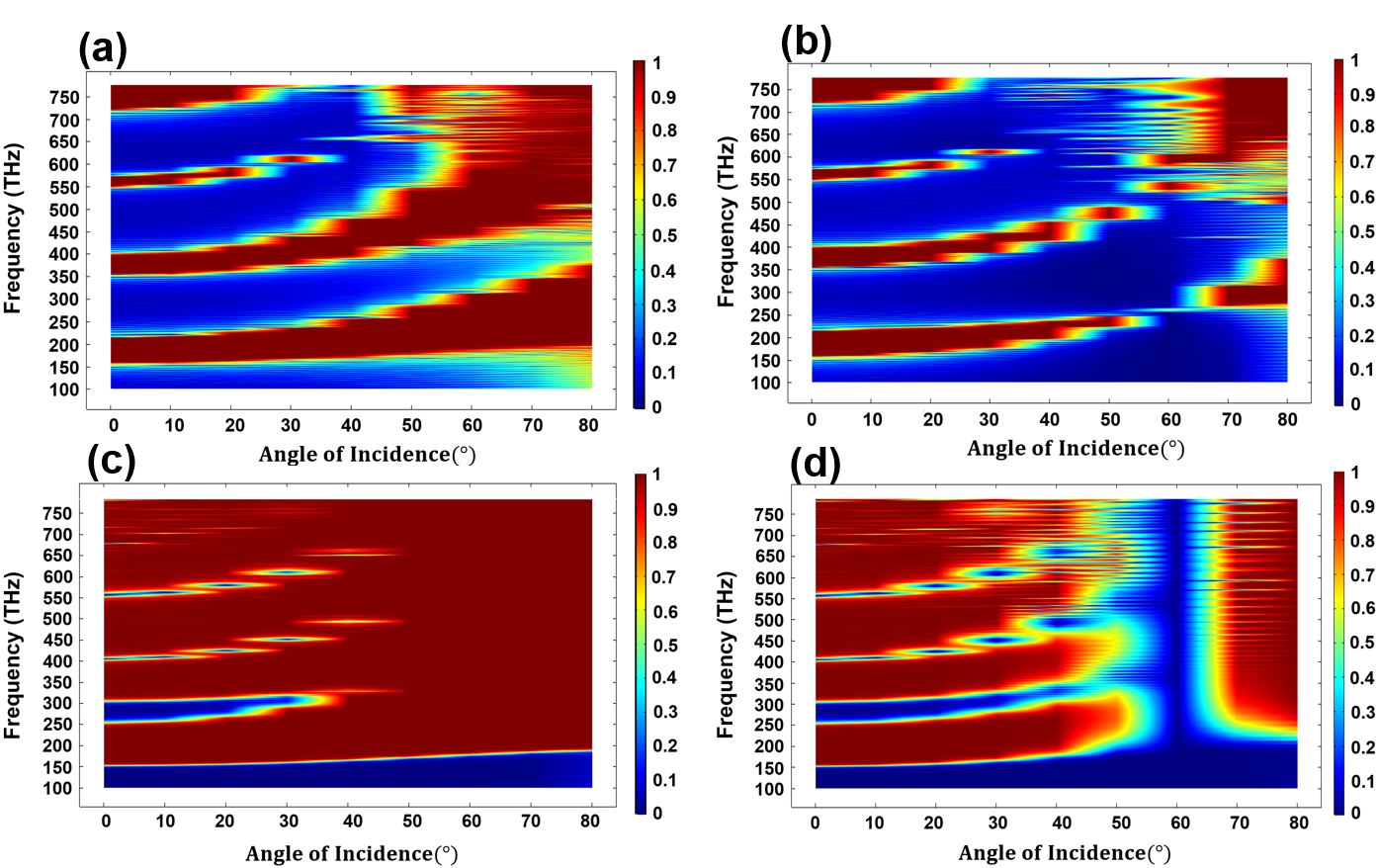}
\caption{a) and b) shows the variation reflection spectrum for TE and TM polarization respectively in a normal DBR ($\Lambda = 400~nm$ and $d_1 = d_2$) as a function of angle of incidence. c) and d) shows the reflection spectrum for TE and TM polarization respectively in C- DBR as a function of angle of incidence. The C-DBR parameters are $\delta=~10~nm$ and $d_1 = 10~nm$. All the cases, total number of units is $N=39$}.
\label{Figure6}
\end{figure}

\subsection{Oblique incidence and angular dispersion}
A detailed comparison of reflection spectrum and dispersion for a normal DBR and a C-DBR (with $\delta = 10~nm$) is elucidated in Fig. \ref{Figure6}(a)-(d). The dependence of reflection spectrum on angle-of-incidence (AOI) for TE and TM polarization in a normal-DBR is shown in Fig. {\ref{Figure6}}(a) and (b) respectively. The oblique incidence essentially leads to a smaller value of normal component of the wavevector and consequently, the PBGs shift to higher frequencies with increasing angle of incidence. As expected, the PBGs tend to broaden for the TE polarization on oblique incidence. On the other hand, the low frequency PBGs for TM polarization tends to reduce up to an AOI = $\approx60^\circ$ and increases thereafter. The drop in PBG for TM polarization is essentially due to the Brewster's effect at the interface of high and low-index layers. Figure \ref{Figure6}(c) and (d) represent the reflection spectrum for the TE and TM polarization respectively for C-DBR. A comparison between Fig. \ref{Figure6}(a) and (c) reveals that the C-DBR exhibits appreciably broad PBGs with very narrow transmission bands separating them. At oblique incidence, the PBG in C-DBR broadens further and shifts to higher frequencies. This is accompanied by shrinking of transmission bands. In fact, the PBGs tend to overlap for AOI $\geq~50^\circ$, thereby leading to a high-reflection band extending from $150~THz$ to $750~THz$ ($1600~nm$ band). Such broad PBGs are usually not found in normal DBR, even with very wide refractive index contrast. It is also interesting to note that the reflection spectrum (in Fig. \ref{Figure6}(c)) exhibits three omnidirectional PBGs which are located in $160-250~THz$, $310-400~THz$ and $490-550~THz$ frequency range. The PBG, in this case (for normal as well as oblique incidence) is limited by the material transparency window and could be extended further (on both spectral ends) through suitable choice of materials. A similar comparison between Fig. \ref{Figure6}(b) and (d) depict broadened PBG in case of TM polarization in C-DBR and narrow transmission bands. However, Fig. \ref{Figure6}(d) exhibits a sharp transmission resonance at the Brewster's angle ($\approx~60^\circ$). The PBGs tend to merge thereafter ($\geq~65^\circ$) giving rise to a broad high reflection band. A comparison between Fig. \ref{Figure6}(c) and (d)reveals that the C-DBR could be employed as a broadband $(\approx 150 - 750~THz)$ polarization filter for $\approx~60^\circ$ AOI.    
\section{Conclusion}
In conclusion, we present an approach to understand the propagation characteristics of modes in a DBR using general techniques adopted in a wide variety of systems exhibiting \emph{SU(2)} dynamical symmetry. The coupled-mode equations describing the forward and backward propagating modes in a DBR could be represented in the form of a single \emph{optical} Bloch-equation where the evolution of state-vector depicts the dynamical behaviour of the wave propagation. This provides a platform to draw an analogy with a \emph{two-level} atomic system and consequently, adopt a formalism for adiabatic evolution in DBR based configurations. In order to realize conditions imposed by adiabatic constraints, C-DBR configurations have been investigated in detail. The DBR variants exhibit enhancement of PBG along with varying degree of suppression of sharp transmission resonances in the reflection spectrum. The impact of alteration in physical parameters of the DBR is explored in detail. It is worth pointing out that the C-DBR configuration involves a discernible longitudinal variation in mode-coupling coefficient $\kappa$ in addition to the sweep in $\Delta k$ (or $\Delta \beta$). Conventionally, such adiabatic process have a close resemblance with the \emph{Allen-Eberly} scheme defined in the context of population-transfer in \emph{two-level} atomic systems\cite{eberly}. Novel DBR designs could further be explored which satisfy the adiabatic constraints. An interesting extension of this proposal would be to investigate the evolution of \emph{geometric}-phase in such DBR configurations and the possibility to control the backscattered (reflection) phase through suitable DBR design. This promises to provide an unique and flexible platform for tailoring the spatial features of an optical beam using adiabatic DBR configurations. Nevertheless, a natural extension of this proposal would be to explore the viability of this formalism in \emph{two}- and \emph{three}-dimensional photonic crystals with focus on applications such as sensing and enhanced nonlinear optical interactions.

\section{Disclosures}
The authors declare that there are no conflicts of interest related to this article.
\bibliographystyle{unsrt}  
\bibliography{ref}  


\end{document}